\documentclass{PoS}

\title{Microquasars: Progress made and open questions}

\ShortTitle{Microquasars:Open questions}

\author{\speaker{I.F. Mirabel}\thanks{On leave from CEA-Saclay. France}\\
        European Southern Observatory. Alonso de Cordova 3107. Santiago, CHILE\\
        E-mail: \email{fmirabel@eso.org}}

\abstract{In the last talk of the conference I summarized the 
main progress and contributions to high energy astrophysics made by studies of microquasars 
in our Galaxy. To stimulate the general 
discussion I have underlined some of the questions that will guide in the near future the research 
in this area of astrophysics. Here I present the viewgraphs and questions formulated during the 
general discussion.}

\FullConference{VI Microquasar Workshop: Microquasars and Beyond\\
		September 18-22 2006\\
		Societ\`a del Casino, Como, Italy}

\begin{document}

\section{The microquasar 1E 1740.7-2942 still a mystery}

Figure 1 shows the Chandra and INTEGRAL images of the Galactic Center region, 
where the most luminous source in the hard X-rays is the famous Einstein source  
1E 1740.7-2942, which is a source of relativistic jets that extend up to a few light years.
The flux, spectrum shape and time variability in the X-rays are similar to those that 
would be observed from Cygnus X-1 located in the Galactic Center region, which supports the 
hypothesis that 1E 1740.7-2942 is a stellar-mass black hole binary.    
However, until today it has not been found -even using the VLT in the near and mid-infrared-, 
an unambiguous optical or infrared counterpart of the donor star in this putative black hole binary.

\begin{figure}
\includegraphics[width=1\textwidth]{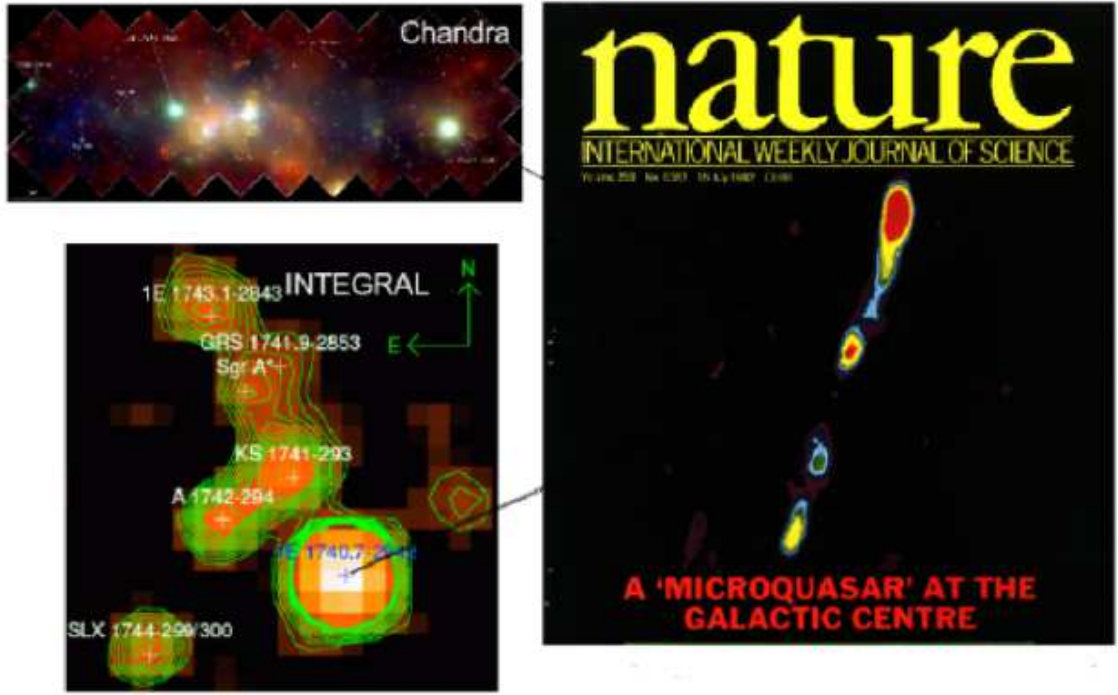}
\caption{Chandra and INTEGRAL images of the Galactic Center region on the left. On the right is 
shown the $\lambda$ 6cm compact counterpart of the high energy source at the center of collimated 
jets that are the trace of synchrotron electrons and possible positrons streaming away from the 
compact source up to distance of a few light years.}
\label{fig1}
\end{figure}

\section{Gamm-rays from compact binaries: microquasar jets or pulsar winds ?}
 
Recent observations have shown that some compact stellar binaries radiate the
highest energy light in the universe. The challenge has been to determine the
nature of the compact object and whether the very high energy gamma-rays are
ultimately powered by pulsar winds or relativistic jets (Figure 2).

\begin{figure}
\includegraphics[width=1\textwidth]{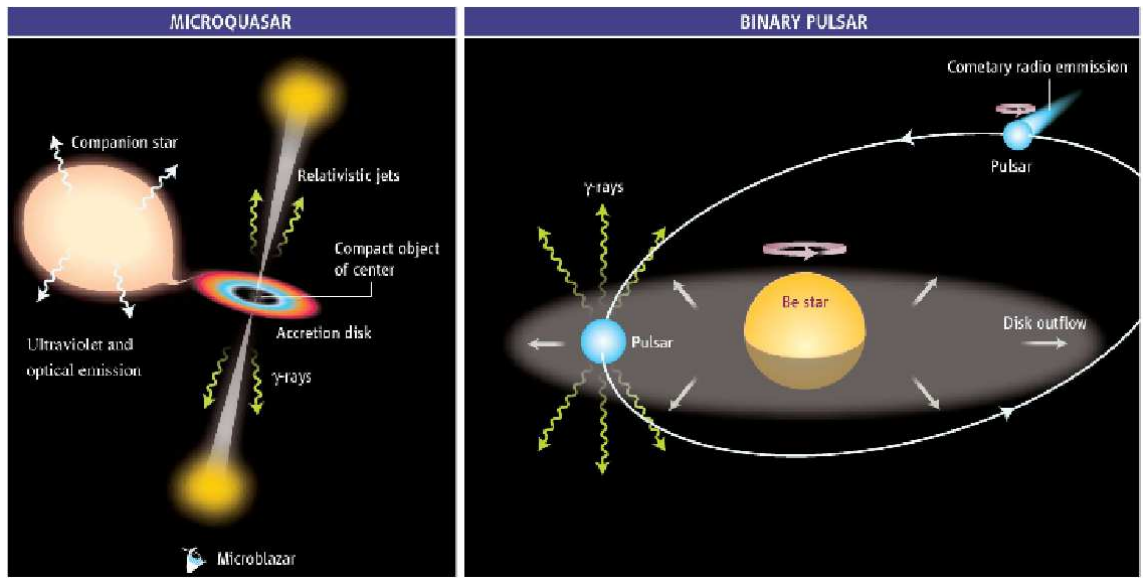}
\caption{Alternative models for very energetic gamma-ray binaries.
{\it Left}: Microquasars are powered by compact objects (neutron stars or
stellar-mass black holes) via mass accretion from a companion star. The jets
boost the energy of stellar winds to the range of very energetic gamma-rays. 
{\it Right}: Pulsar winds are powered by rotation of neutron stars; the wind
flows away to large distances in a comet-shape tail, as has been shown in
to be the case for LS~I~+61~303. 
Interaction of this wind with the companion-star outflow may produce very energetic gamma-rays.}
\label{fig2}
\end{figure}

Multiwavelength
observations have shown that two (PSR~B1259$-$63 and LS~I~+61~303) of the three
gamma-ray binaries known so far are neutron star binaries and that the very
energetic gamma-rays from these two sources may be produced by the interaction
of pulsar winds with the wind from the companion star. In fact, in this meeting 
were presented observations at radio
wavelengths that support the idea that LS~I~+61~303 may be a gamma-ray
pulsar rather than a microquasar. As expected from the pulsar
wind model (right panel of Fig. 2), VLBA images of the radio
emission show a relativistic wind from the compact object that spins as a
function of the orbital phase. However, no pulsations have so far been detected 
from LS~I~+61~303. 

At this time it is an
open question whether the third gamma-ray source (LS~5039) is also powered by a
pulsar wind or a microquasar jet, where relativistic particles in collimated
jets would boost the energy of the wind from the stellar companion to TeV
energies. High resolution maps at radio wavelengths would tell whether the direction 
of the jets spin as a function of the orbital phase or their position angle in the sky 
remain the same. On the other hand, it would be interesting a search for TeV emission 
from confirmed black hole binaries, which would provide support to the microquasar 
jet mechanism for the production of very energetic gamma-rays. 

These TeV binaries do not require Doppler boosting, namely, they are not microblazars.

\section{Perspectives to determine the spin of lack holes}

Three methods have been proposed to determine the spin 
of black holes. They are based on measurements of: 
1) the temperature of the continuum spectrum. Due to the shrink 
of the inner accretion disk in rapidly rotating (Kerr) black holes, the 
temperature increases with angular momentum.
2) the Fe line at 6.7 keV, which is skewed towards lower energies as a function of the 
angular momentum.  
3) the maximum fix frequency of quasi-periodic oscillations  (QPOs), which is  
a function of the basic parameters that define astrophysical black holes, namely, the 
mass and the spin. 

In this meeting new exciting insights on the spin of black holes were presented. 
Refining the analysis of the continuum spectrum it was found that GRS 1915+105 contains a 
black hole spinning near the maximum rate, whereas GRO J1655-40, LMC X-3, and 4U 1543-47 
have moderate spins. It is interesting that the source with maximum spin is also one of 
the most powerful jet sources. However, due to the small number statistics it is still 
an open question whether the power of relativistic jets is a function of the spin, as 
predicted by different models.  

In this field there are still the following basic caveats:

a) The physics of QPOs is still not fully understood. In this meeting it has been proposed 
that they are due to global modes that arise as spontaneous instabilities rather than 
as orbiting ``blobs''. 

b) It is still uncertain how the forest of warm absorber lines may affect the X-ray continuum, 
and therefore the estimates of the temperature of the inner accretion disk. 

\section{Dark outflows and super-Eddington sources}

In last years there have been increasing evidences showing that relativistic and non-relativistic 
mass outflows from accreting compact objects may be radiatively inefficient. The evidences  
reside mostly in the interaction of the jets with the interstellar environment. They are: 
1) The moving 
X-ray jets from the microquasars XTE J1550-564 and H1743-322 that produced in real time synchrotron 
bow shocks far away from the sources and long after X-ray outbursts from the compact binary, and 
2) The large scale interstellar structures inflated by the relativistic jets as observed at 
radio continuum wavelengths 
in Cygnus X-1 and SS 433 (Figure 3).

\begin{figure}
\includegraphics[width=1\textwidth]{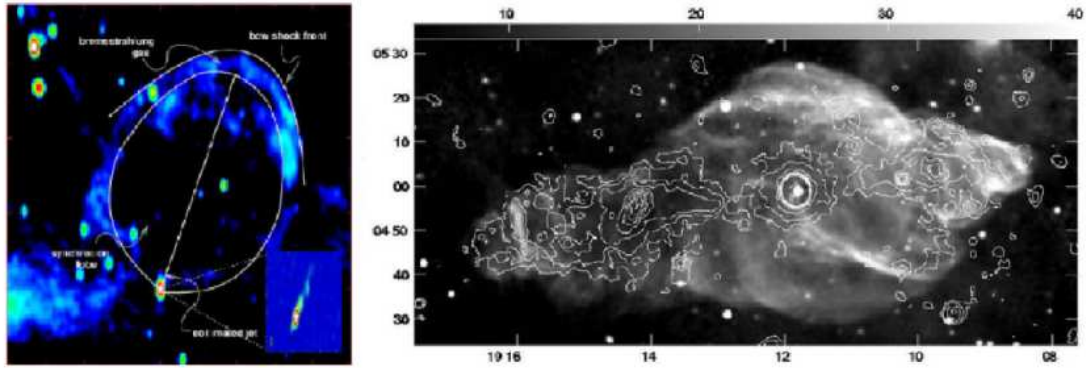}
\caption{Left panel: Jet-powered nebula around Cygnus X-1 from Gallo et al.(2005).
Right panel: Image at $\lambda$20cm of the nebula W50 that hosts SS 433, from Dubner et al. (1998).}
\label{fig3}
\end{figure}

In SS 433 a  mechanical luminosity of the relativistic jets of 
$\geq$ 10$^{39}$ erg s$^{-1}$ has been estimated from spectral lines of ions of H, He, and Fe at 
optical, infrared and X-ray wavelengths. In SS 433 more than 50\% of the energy is not radiated 
and it has been estimated that more than 30\% of the accreted mass is being ejected in the form 
of massive 
winds and relativistic jets. 
Clearly, SS 433 is a super-Eddington source of stellar mass with anysotropic outflow and 
radiation that from 
other viewing angle would be classified as an ultraluminous X-ray source (ULX). It may be used 
as a probe of ULX's in external galaxies. 

The following questions remain open:

1) What is the relation between massive winds and warm absorbers ?

2) If black holes of intermediate mass ($\geq$ 100 M$_{\odot}$) are prolific, why no one has been 
identified in the Milky Way or the Magellanic Clouds, where the dynamical mass of compact objects 
in binaries can be determined ?

\section{Disk-jet coupling}

In 1994 it was observed a mayor ejection event from GRS 1915+105 that was followed 
simultaneously with BATSE on the Compton Observatory and the Very Large Array (Figure 4,
left panel). The connection between the inner disk-corona and the jets became  
clear because the ejection took place at the time of a sudden drop in the 20-100 keV flux.
At that time it was assumed that the high energy electrons producing the hard X-ray radiation 
had been suddenly blown away in the form of collimated relativistic jets. The physical nature 
of the instability that produced such transition remains unknown. 

\begin{figure}
\includegraphics[width=1\textwidth]{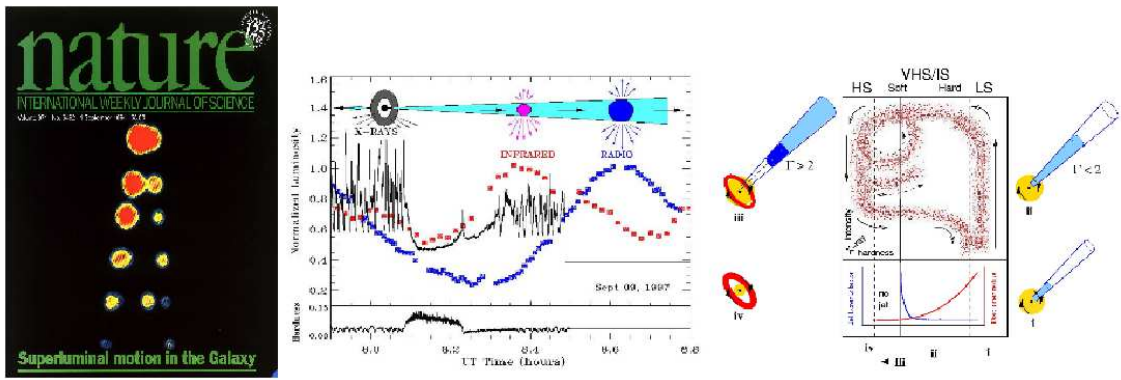}
\caption{left panel: Superluminal ejection in GRS 1915+105; central panel: multiwavelength
simultaneous observations of instabilities in the accretion disk and the genesis of jets;
Right panel: Unified model for black hole accretion-jet phenomena (Fender et al 2004).}
\label{fig4}
\end{figure}

Four years later was obtain further insight into the connection 
between accretion disk instabilities and the formation of jets.  
In $\sim$1 hour of simultaneous multiwavelength observations of GRS 1915+105 
during the frequently observed 30-40 min X-ray oscillations in this 
source, the connection between sudden drops of the x-ray flux from the accretion disk 
and the onset of jets was observed again in several occasions (Figure 4, central panel).
The triggers of jets are instabilities in the accretion disk 
(transition from low-hard to high-soft states). The X-ray spike in the central panel marks 
the onset of a shock through a compact, steady jet. 

A comprehensive phenomenological model based on the notion that the jet phenomenology is part 
of the accretion process was later proposed by Fender et al (2004; Figure 4 right panel), 
to which I refer the reader for a detailed description.  Analogous transitions have now been 
observed in the quasars 3C120, 3C279 and 3C390, the diagram in the right panel of Figure 4 
being considered in observations of the disk-jet coupling in accreting black holes of all 
mass scales.
 
\section{The black hole at the Galactic Center}

European and American teams have been studying the kinematics of the central cluster of massive 
stars with the VLT and Keck respectively, where using adaptive optics at infrared wavelengths 
from the ground becomes competitive with space astronomy. Both teams reach similar results  
and estimate a black hole mass of 3 $\times$ 10$^6$ M$_{\odot}$. However, there are 
following puzzles:

1) How could a massive cluster of massive stars be formed and survive in a region of 
10 light years radius from the supermassive black hole ? 

2) Observing with the VLT, QPOs of about 15 min were reported. However, no such QPOs have 
been observed so far with 
Keck. In microquasars it is known that QPOs of maximum fix frequency are only  observed 
sporadically but not permanently. Could the presence and absence of these QPOs be due 
to chance ?      

3) The flares in Sgr A$^*$ exhibit similar time delays between the X-rays, infrared 
and radio wavelengths as observed in the flares of GRS 1915+105 (Figure 4 central panel). 
These time delays are comparable, irrespective of the mass of the black holes which  
indicates that the radiating plasma has detached from the gravitational field of the black hole.

\section{Microquasars and beyond}

In the three astrophysical manifestations of astrophysical black holes are found the same three 
basic ingredients: a black hole, an accretion disk, and relativistic jets (Figure 5).

\begin{figure}
\includegraphics[width=1\textwidth]{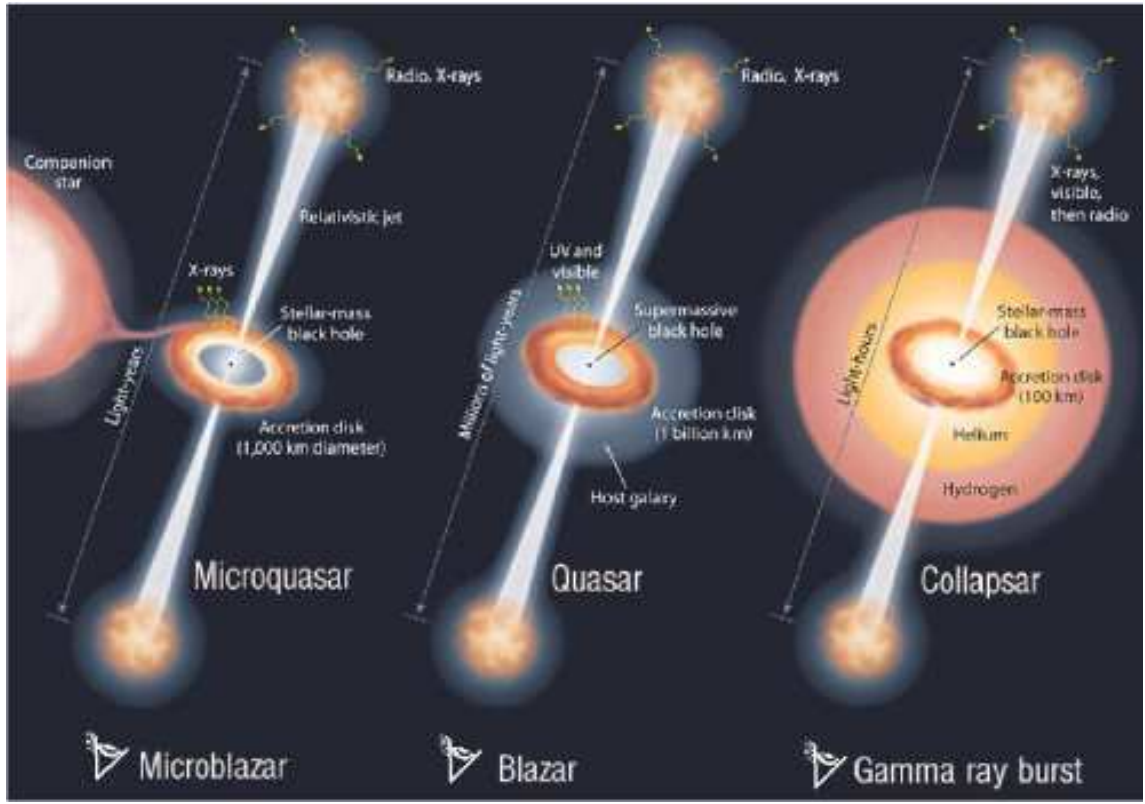}
\caption{The three manifestations of astrophysical black holes.}
\label{fig5}
\end{figure}

For microquasars, quasars and GRBs have been proposed the same alternative models:

Jet models: Discontinuous ejections of blobs (internal shock model) versus pointing flux.

Two flow model: collimated MHD flow plus super-relativistic pair beamed jet

Jet production models: MHD centripetal (inner) plus magnetic dissipation at larger scales.

 \section{Microquasar - GRB connections}

In previous publications I had proposed that  microquasars in our Galaxy 
can be used as nearby probes of the collapsar model for the  formation of stellar-mass black holes. 
In particular, the kinematics of microquasars can be used to constrain the energy of natal 
explosions during the formation of black holes. For instance, from the chemical composition 
found in 
the atmosphere of the donor star and runaway velocity of GRO J1655-40 it was 
proposed that the black hole in this microquasar was formed through a very energetic supernova. 

The question on whether all black holes form through energetic explosions or some form by 
direct collapse is related to the question on whether all long GRBs are associated with 
hypernovae (type Ib/c). The recent observations of GRB 060614 by Della Valle et al. (2006) show 
that some long-lasting GRBs can be associated with very faint supernova. This result is consistent 
with the finding that the black holes of $\geq$ 10 M$_{\odot}$ in Cygnus X-1 and GRS 1915+105 
have not receive a large linear momentum during formation (Figure 6).

\begin{figure}
\includegraphics[width=1\textwidth]{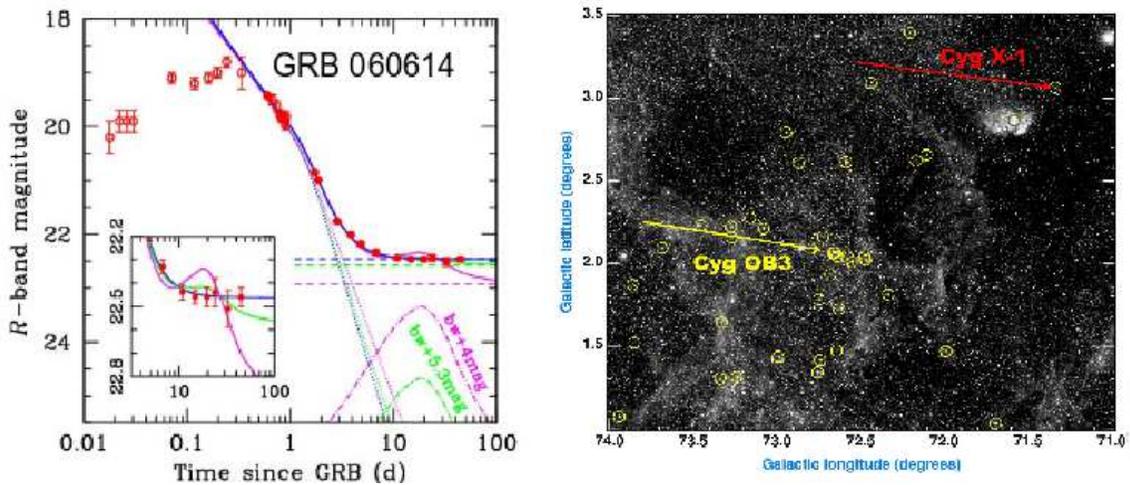}
\caption{Left panel: R-band light curve of the GRB 060614 afterglow showing no
luminous supernova (Della Valle et al. 2006). Right panel: Proper motion of of Cygnus X-1
and the parent association of massive stars Cygnus OB3 showing that the microquasar remained
at its birth place and did not received a strong kick.}
\label{fig6}
\end{figure}

 \section{Empirical correlations and analogies between stellar and black hole astrophysics}

Different empirical correlations between the X-ray and radio fluxes, 
noise spectrum, and QPOs as a function of black hole mass have been proposed. 
If these empirical correlations 
become more robust, independently of the models, the mass and spin of black holes will 
be determined. As in stellar astrophysics, in black hole astrophysics the luminosity is a 
function of size and temperature. 

At present, black hole astrophysics is in an analogous situation as was
stellar astrophysics in the first decades of the XX century. At that
time, well before reaching the physical understanding of the interior of
stars and the way by which they produce and radiate their energy,
empirical correlations such as the HR diagram were found and used to
derive fundamental properties of the stars, such as their mass.
Analogous approaches are taking place in black hole astrophysics. Using
correlations among observables such as the radiated fluxes in x-rays and
radio waves, quasi-periodic oscillations, flickering frequencies, etc,
fundamental parameters that describe astrophysical black holes such as
the mass and spin of the black holes are being derived.

\end{document}